# A Fully Distributed, Privacy Respecting Approach for Back-tracking of Potentially Infectious Contacts

Adam Wolisz
Professor emeritus, TU Berlin; Fellow, ECDF Berlin; awo@ieee.org

**Abstract**: In limiting the rapid spread of highly infectious diseases like Covid-19 means to immediately identify individuals who had been in contact with a newly diagnosed infected person have proven to be important. Such potential victims can go into quarantine until tested thus constraining further spread. This note describes a concept of mobile device (e.g. Smart phones) based approach for tracking inter- personal contacts which might have led to infection and alerting the potential victims. The approach assures means for defense against malicious usage while assuring a high level of privacy for all people involved.

## 1. Introduction

The pandemic spread of actual COVID-19 is mainly due to the fact, that infected individuals are spreading the virus long before they expose any symptoms. It is pretty obvious that after a person has been diagnosed infected, it is beneficial to immediately ask his previous contacts – potential victims - to self- quarantine until reliably tested. In this way the potential victims do not endanger further persons. Due to their wide usage smart of phones it is tempting to use them to discover potential victims. The experience of several countries, notably China and South Korea, in fighting the outbreak has demonstrated the usefulness of such approach.

Tracking people movement using the location features of smart phones is by far not a new idea. It goes back to the Cambridge FluPhone [1]. China and in South Korea use the location data from the mobile communication providers (telcos) to identify the previous contacts. For this purpose, the whole trajectories of each person have to be stored (at least for some period, usually 2- 3 weeks) and in order to enable proper post processing and detection of periods of "past closeness" to the trajectories of a person diagnosed sick. This approach triggers, however, in multiple societies justified concerns about the far- reaching privacy violation, in some countries such means are even forbidden by law. Approaches which might remove – or at least substantially limit the privacy violation, while offering at least equivalent identification of the past contacts are definitely needed!

In answer to these concerns very recently the MIT researchers [2] have published a concept in which the trajectory of the movement is stored locally on everybody´s mobile device, and can be uploaded to a trusted cloud. A public service associated with data stored at the public cloud would make it possible to anybody to check if his trajectory has a commonality with any of the already stored trajectories of infected person. The work on that is in progress [3].

Such approach avoids the general pre- storing of movement trajectories of all, and only the trajectories (probably pre-edited by the owner!) of the person diagnosed sick are centrally stored. The authors admit, however – that there is a possibility of "fake information" about trajectories of sick persons to be uploaded, in order to create false suspicions of infected areas (e.g. businesses). Some more comments can be found in [4]. According to recent press/TV information the idea of a fully distributed solution based on Bluetooth seems to appeal to numerous (see e.g. [5]).

In this paper a Bluetooth based approach, respecting the privacy of all actors involved and assuring reasonable security against malicious usage is presented. In Section 2 we present the fundamentals of the approach. In Section 3 some partial services and implementation issues of the suggested solution are discussed. In Section 4 the features of this approach are summarized, and a strategy for step-wise deployment and possible extensions is presented.

## 2. Outline of the approach -the basic version

We assume that everybody is in possession of a mobile device (smart phone, intelligent wearable) enabling download of new applications and storing some data and equipped with a Bluetooth interface. In addition, we expect the device to support the notion of time and – some notion of location (a high precision of the location service is NOT required!)

We assume that each person will generate a "Pseudo- ID"- further called PID. The PID could be any rare (preferably locally unique) sequence of symbols which will be used exclusively for the sake of tracking the disease spread. Within this approach there is no need whatever to map the PID back to a real person. The PID can be changed in at any point of time - simplicity we will initially assume that a single PID is used a given individual within one spread of a given infection (e.g. one season). In addition, each person should make available an reachable address for notifications called Pseudo- Address, shortly PAD).. An example of a PAD might be an e- mail address like PID@xxx.com with any e-mail service XXX. While a change of the PAD is possible at any time! – for simplicity of the explanation we will initially assume that the PAD is also fixed for a given user over one spread of the given infection.

After generation of the PID and PAD and the KEY a routine operation – collecting the contacts follows:

### 2.1. Collecting the contacts

Each device should have the Bluetooth interface permanently enabled, and running a *Contact identification* service. In essence this service should have the following functionality:

- discover another Bluetooth device.

- exchange with that device the information records (post own, receive that of the peer). The information record consists of the following fields: The PID, the PAD, local time and local position – both time and position from the point of view of the transmitting device.

- Decide if the contact has been close enough and long enough (additional criteria might follow!) to be classified as a *Significant Contact.* If so, store a "log entry" consisting of the content of the own transmitted record and the content of the record received from the peer. A contact is considered a significant one, if it fulfills pre-defined criteria as for the mutual distance of the devices and duration of staying within this distance. For contacts not decided to be significant, the received data are deleted. Each log entry should be stored for the time period recommended by the epidemiologists – for the COVID-19 being 2- 3 weeks.

Note that there is no requirement of tight time synchronization, the "time" nor is there a need to use the same abstractions as for the location by parties involved in a contact. As the matter of fact only the PAD has to obey the usual standard format for a usable address, the remaining fields are just sequences of symbols which will never be interpreted.

The log can be stored at the mobile device of the user, or in some private storage belonging to him in a cloud- presumably encrypted by the user's individual encryption scheme. Under no circumstances is a third party expected to access this log.

**2.2. Actions upon infection**

If anybody participating in the contact tracing according to this concept becomes diagnosed infected the following three steps should be executed:

A/ Certificate of Infection

All labs/doctors diagnosing equipped/entitled to diagnose the infection should possess a feasibility to use some widely accepted asymmetric cryptography, with public key being easily accessible. Ideally the list of such units and their public key should be run in a well-known directory on a set of reliably maintained server.

Upon diagnosing that an individual is infected the lab/doctor should generate a simple "certificate of infection" featuring the date of the test and identifying the infected person only by his PID (or list of PIDs) used in last relevant period (e.g. 2-3 weeks) as provided by the infected individual. Some information on the estimated time in which the diagnosed person could have already been infectious to others could be added. This electronic certificate (in some generally readable format, e.g. pdf) should be encrypted with the private key of the issuing lab/doctor and posted to the PAD od the infected individual.

B/ The notification service
A notification service running on behalf of a user (on his mobile device or wherever!) and having access to his log should post to PADs of all the significant contacts in the log the warning of possible infection. This notification should contain:

   The PID used by the infected individual, as well as the time and location received from the peer during the contact (as retrieved from the log), plus the medical certificate of infection with indication of its encryption with the personal key of the lab/doctor. There is no requirement that the notification be posted from the service connected to the PAD of the infected user!

2.3. **Receiving and validating the notification**

Any user should keep access to his information service associated with the PAD (e.g. his e-mail!) and check regularly the information incoming through this PAD. A verification service processing the notification should be activated on reception of a notification.

By comparing with the own log, the verification service can cross check that there has been, indeed at the claimed time and location a contact with a device announcing the claimed PID. In such way fake claims of contact can be filtered out. Afterwards the verification service should decrypt the certificate of infection (using the trusted source for getting the public key of the lab/doctor) and cross check the PID in this certificate against the PID in the notification (and own log) in order to verify the fact of infection.

After completion of these verifications the user obtaining the notification can safely assume its correctness and start acting appropriately (going in self quarantine, consulting his doctor, requesting a test, etc.

**3. Comments on partial functions of the basic design and their implementation**

The concept presented above can be easily mapped into an application loadable to the mobile device, easily structured in a set of interacting services, plus a separate service for certificate generation. Such service-oriented structure would allow to exchange some functions (versioning the application) – e.g. by improving the decision on which contacts are significant, or adding additional security precautions (e.g. as commented in section 4). Note that the applications following this concept could be executed entirely on local devices –strengthening the feeling the privacy.

Let us now comment a bit more about some selected parts of the functionality.

3.1. **Generation of PIDs and PADs**

We have assumed that the PIDs are temporarily constant during a single disease spread. This assumption is not necessary, in fact the PIDs could be modified any time. The only important constraint is, that ALL the relevant PID should be declared to be introduced in the Infection certificate! In view of the richness of exchanged data a false positive in indicating a significant contact due to limited multiple use of the same PID by different persons has a low probability (a possibility of increasing the robustness of the presented approach requiring a specific generation of the PIDs is explained in section 4).

Similarly, there are no specific limitations as for usage of different PADs – the only requirement is, that the user keeps access to these PADs for the required period (likely 2- 3 weeks).

3.2. **Discovery of other Bluetooth devices, data exchange and distance estimation.**

This question has been widely discussed in the literature, beginning with the classical paper [6], with relevant further improvements discussed in [7], [8] and recent [9],[10] and references therein.  This functionality should be implemented as a separate service featuring a protocol which precise specification reminds to be chosen.

Similarly, there exists a large literature on distance estimation among Bluetooth equipped devices. This is not a trivial issue, as the pure signal strength might be seriously despared e.g. by shadowing the device form its "peer" carried by another person by the bodies of their owners. Here also exists a huge literature – beginning with the classical paper [11] and interesting contributions in [12] and [13] claiming a respectful accuracy. A very attractive method of device distance estimation by combining Bluetooth with acoustic signals generated/received by the smart phones in frequencies not audible to humans has been described in [14]. This approach eliminates the adverse effect of the shadowing by the human body.

Again – distance estimation should be encapsulated as a separate service, which could be improved in consecutive versions.

3.3. **Identification of significant contacts**

Which criteria are to be used for deciding if being relatively close has created a risk of infection is definitely an open issue which, however, has anyway to be solved by any approach – notably also in case of mobile data tracking by telecom service providers or usage of GPS (which, btw. Is constrained to outdoor scenarios). At first glance it seems that the distance and duration of staying in a given distance seem to be primary criteria, while additional dependence on environmental conditions (in-door vs outdoor, temperature, coughing, etc. might be of significance.

Let us note here that usage of smart phones opens the potential to use additional sensors (e.g. recognize coughing or sneezing while in proximity of others!). Making decision about the significance of a contact – meaning there is a reasonable probability of infection spread during the contact – should in any case be implemented as a separate service, allowing to replace the decision algorithms as the understanding of the spread of a given virus improves. The strategy to start with "rather pessimistic" criteria (leading to higher rate of "false alarms") or with rather optimistic (in which case "almost sure" infections are indicated, but further could be ignored) is a hard decision to be made…

**3.4. The notion of location and time announced to the peer in contact.**

In this approach the real proximity of the devices is NOT related to the location declared by a given device but derived only due to the local signal exchange. The notion of location as exchanged in the information records is used exclusively to validate the claim of having been involved in a mutually significant contact as indicated by the received notification (he knows what I have called location at the time of the contact!). Therefore, the locations could be represented by semantic notions like: on the walk, during exercising, at the social meeting which meaning is entirely private to the given user and meaningless to anybody else! This strengthens additionally the privacy of location, being well known as intrinsically difficult to preserve!

As for time – we have already mentioned that there is no need to keep a tight time synchronization at participating devices. The notion of the local time of the contact should be precise enough to be used – jointly with the date of test listed in the certificate – in order to decide about the point in time where a test might be useful, and length of the quarantine. In addition, it is used – jointly with the information on location – to verify the credibility of the contact claim.

**4. Deployment Strategy and Envisioned Extensions**

In previous section we have pointed out several design options and in some cases a need for further research. On the other hand, the urgency created by the spread of the COVID-19 justifies quick deployment and usage of any, socially accepted technology which might help in mitigation of this spread. The presented approach allows for a step-wise release strategy, in which first a "quick and semi-dirty" but useful solution might be released, and slowly replaced by improved versions. While the basic service oriented- software architecture and the basic mechanism for data exchange establishment among the "close" Bluetooth devices should be set from the very beginning, multiple other algorithms can be initially implemented in a rather simplistic form and gradually replaced in later updates. Out of experience one must, however, consider that not all users will always have the newest version, so precautions for "interoperability" between the older and newer versions have to be taken. Let us present an example of such analysis for the algorithm making the decision if a contact is significant for spreading the infection on basis of the relative distance.

Just for the sake of explanation let us assume that an initial version assumes the distance under 3 meters to be a risk and this version is installed on the device X, why another device Y has a newer version limiting the critical distance to 1,5 meters. What will happen in case of a contact with distance 2,25 meters in such case? The device X will consider the contact significant, while device Y will ignore it and not introduce any entry in the log. If the owner of the device X is diagnosed sick, a proper notification will be posted, but ignored by the device Y as a fake (no matching entry in the local log). If the owner of the device Y is diagnosed sick, no notification will be posted to the device Y because there is no information about a proper contact! In both cases the newer strategy is prevailing. Similar analysis is possible in case of gradually introducing better distance estimation methods. Other cases need separate consideration along the same line.

For the sake of quick deployment initial deletion of the "certificate of infection" issuing might also be considered. This mechanism has been introduced to avoid "Jokes": sending notification by people who maliciously claim sickness. Note that such claims could be only considered as valid, if the sender credibly presents a PID which the receiving party has in her log at the time and location claimed by the sender! In this approach such malicious claims could be misleading only if the "joking person" really made a "contact" with those to be "wrongly alerted". The probability of such action – especially if everybody tries to obey social distancing is low. Thus – at least in periods in which bigger crowds are prohibited – the "certificate of infection" might not be necessary. This would significantly simplify the deployment – as the specific service supporting issuing such certificates might be ignored. Assuming proper format of the notification – such certificate might be introduced later, if needed.

On the other hand, the basic mechanism does not prevent people from forging the notification of contact by copying such notification obtained by – say – a friend, and replacing his PID by the own PID. The reason for such action might be an attempt to get tested in case of limited testing capacities and health authorities giving precedence to those providing some "evidence" of contact with the infected persons. This might become an issue especially in case of serious bottlenecks in testing capacity. One straightforward option to verify such claims is the combination of two measures:

- Creating a repository of the "notified PIDs" fed with data while issuing the "certificate of Infection" and accessible for those authorized to make decision about testing a person seeking such test.
- Verification of ownership of the entry on the repository of "Notified PIDs" by formalizing the process of PID generation. For this purpose, the deployed application should be extended by a "trusted PID" generator, following procedure: deriving PIDs – at any time! – as a one-way hash on the record containing [*personal data of the user, any arbitrary phrase invented by a user*]. In case of doubt if the PID in a notification of contact with a person diagnosed sick (as proven by the certificate od infection!) has in the past REALLY been used by the claiming individual, he might be requested to prove it, by proving the knowledge of the proper phrase. Adding some personal data (easily verifiable and widely known, like first and family name) to the record used for the hash generation – as mentioned above – would prohibit "just sharing the phrase"!

Finally, let us also indicate, that this concept could be also extended toward the defense of business owners against fake claims of visits by people tested later as infected. For this purpose, a version of application with a certified non- changeable log (e.g. following the principle described in [15]) could be created, and customers requested to approve that the log containing their PIDs collected while patronizing the place might be published if necessary for such defense (as PIDs can be changed any time this would not be a real privacy leakage of third parties). Entries in such log – jointly with a query in the above-mentioned repository of "Notified PIDs" could provide an evidence of possible false claims (the claiming individual did not visit the business, or is not certified sick).

## 5. Final Comments.

In this paper a concept of fully distributed, privacy respecting approach to back- trace contacts causing possible infection has been presented. The user is, at any time, in full possession of her data, and there is no obligation whatever to share any of the data. Devices used for recognition of the significant contacts according to this concept do NOT have to be continuously connected to the internet – which makes the described approach robust to possible connectivity gaps!

In addition, the basic design extended by the above discussed extensions seems to offer a rather strong security against malicious usage. A strategy for step-wise deployment enabling a rather quick start and gradual improvement has been also explained. Up to the best knowledge of the author this is the first description of a systems fulfilling all the above claims.

Let us point out, that in addition to the warning about a potential infection, usage of a system like the described above, might help the user in increasing her safety. Study of the own log allows to recognize the frequency of "close, potentially dangerous encounters" – and also discover locations in which one is most exposed to such encounters. This itself, might lead to improved social distancing and thus significantly contribute to reduction of the infection spread!

It is assumed that participation in the tracking of contacts – as described above – should be based on voluntary decisions. Obviously, some of the users might decide to us rather direct definition of locations (or the explicit position) and donate the log contents for statistical analysis by researchers.

While back- tracking of the potentially infectious contacts using smart phones is very attractive, one should not forget, that assuring a continuous operation of a single application on smart phone platform is not trivial itself. For indicating the challenges and already discovered problems indicate the recent studies [16] and [17]. The solution discussed in this paper has been presented in the context of smart phones, but in fact it is feasible to extend its usage also towards other programmable wearables- some of them might be more robust in the sense of vulnerabilities mentioned in the above references.